# Title

**Millimeter-scale topography enables coral larval settlement in wave-driven oscillatory flow**

## Authors and affiliations

Mark A. Levenstein,[1,2,†] Daniel J. Gysbers,[3] Kristen L. Marhaver,[4*] Sameh Kattom,[1] Lucas Tichy,[4,5] Zachary Quinlan,[6] Haley M. Tholen,[1] Linda Wegley Kelly,[6] Mark J. A. Vermeij,[4,7] Amy J. Wagoner Johnson,[1,2,8*] and Gabriel Juarez[1*]

[1] Department of Mechanical Science and Engineering, University of Illinois at Urbana-Champaign, Urbana, IL 61801, US,
[2] Institute for Genomic Biology, University of Illinois at Urbana-Champaign, Urbana, IL 61801, US,
[3] Department of Physics, University of Illinois at Urbana-Champaign, Urbana, IL 61801, US,
[4] CARMABI Foundation, Piscaderabaai z/n, PO Box 2090, Willemstad, CW,
[5] Department of Microbiology, Rabound University, 6525 XZ Nijmegen, NL,
[6] Scripps Institution of Oceanography, University of California, San Diego, La Jolla, CA 92037, US,
[7] Department of Freshwater and Marine Ecology, Institute for Biodiversity and Ecosystem Dynamics, University of Amsterdam, 1098 XH, Amsterdam, NL
[8] Carle Illinois College of Medicine, University of Illinois at Urbana-Champaign, Urbana, IL 61801, US

[†] *Present address*: Université Paris-Saclay, CEA, CNRS, NIMBE, 91191, Gif-sur-Yvette, FR

M.A.L. and D.J.G. contributed equally to this work

**\*Co-Corresponding Authors:**
    Kristen L. Marhaver; kristen@marhaverlab.com
    Amy J. Wagoner Johnson; ajwj@illinois.edu
    Gabriel Juarez; gjuarez@illinois.edu

**Abstract**

Larval settlement in wave-dominated, nearshore environments is the most critical life stage for a vast array of marine invertebrates, yet it is poorly understood and virtually impossible to observe *in situ*. Using a custom-built flume tank that mimics the oscillatory fluid flow over a shallow coral reef, we show that millimeter-scale benthic topography increases the settlement of slow-swimming coral larvae by an order of magnitude relative to flat substrates. Particle tracking velocimetry of flow fields revealed that millimeter-scale ridges introduced regions of flow recirculation that redirected larvae toward the substrate surface and decreased the local fluid speed, effectively increasing the window of time for larvae to settle. In agreement with experiments, computational fluid dynamics modeling and agent-based larval simulations also showed significantly higher settlement on ridged substrates. These findings highlight how physics-based substrate design can create new opportunities to increase larval recruitment for ecosystem restoration.

**Introduction**

The recruitment of pelagic larvae is a critical step in the life cycle of many sessile marine invertebrates, and for reef-building corals, the pervasive failure of natural recruitment processes over the past 50 years has compounded global coral reef losses [1]–[3]. During this critical life stage, swimming larvae can spend days to weeks in the water column and be transported hundreds of meters to kilometers before encountering a suitable location in which to attach and settle [4]. However, because larval dispersal and settlement take place over vast spatial and temporal scales, it is extremely difficult for researchers to observe these processes as they happen in nature; larval dispersal and settlement are consequently referred to as a "black box" [5], [6]. In turn, this knowledge gap impedes the restoration of marine ecosystems, such as coral reefs, through substrate and habitat engineering [7]–[9].

The effectiveness of coral restoration methods relies on mimicking or improving upon natural reef environments and their properties, including the biological and physical factors that aid larval navigation, settlement, and survival. Efforts to understand and promote larval settlement in particular have focused primarily on the biological cues that facilitate the transition from pelagic dispersal to benthic settlement. For example, coral larvae settle in the presence of benthic organisms such as crustose coralline algae (CCA) [10]–[12], bacterial biofilms [13]–[15], and individual organic molecules produced by CCA and bacteria [10], [16]–[19]; these biological cues are now commonly used to promote coral settlement in larval propagation and reef restoration. Several physical settlement cues have also been identified for coral larvae, although these have been less widely studied or applied for coral propagation and restoration. For example, coral larvae swim toward the acoustic signature of a reef [20], avoid areas with relatively high ultraviolet radiation [21], and select settlement locations based on substrate color [22] and substrate topography [23], [24]. Focusing on topography-based settlement preferences, larvae have generally been observed to settle within topographical features close to their size, which are thought to maximize their number of attachment points (i.e., Attachment Point Theory) [24]–[26]. However, this rationale does not fully account for the effects of flow near surfaces, especially when the prevailing hydrodynamic forces are much stronger than larval swimming and adhesion abilities.

Indeed, fluid motion is critical to the maintenance and growth of benthic marine ecosystems. Benthic boundary layer (BBL) flows, i.e., the fluid motion directly above benthic surfaces, influence the exchange of materials and resources between the water column and the benthic zone. For instance, these flows affect nutrient uptake [27], [28], gas exchange [29], [30], surface pH [31], and larval transport and settlement [32], [33]. BBL flows are determined by the fluid-structure interactions between the viscous flow profile of water and the benthic topography [34] and have characteristic fluid regions with velocity gradients (e.g., shear, strain) and recirculation (e.g., rotation). Shallow coral reefs, in particular, experience a range of dynamic BBL conditions influenced by both reef topography and strong unidirectional currents and oscillatory wave action [35]. These fluctuating flow structures produce forces and torques on marine organisms that affect their transport, their settlement, and whether they behave as active or passive particles [36]. In contrast to these dynamic natural environments, many lab-based settlement experiments and tests of restoration substrates are conducted under static or near-static conditions [24], [37]. While these approaches have produced some key developments in restoration ecology, they provide incomplete information about natural recruitment mechanisms and opportunities for further innovation in reef-site interventions.

Here, we demonstrate the potential for millimeter-scale substrate topography to enhance larval settlement by modifying BBL flows. Millimeter-scale ridges produce boundary layers with regions of recirculation and low velocity even in wave-driven flows much faster than larval swimming speeds. In experiments performed using a custom-built, oscillatory flume tank (Fig. 1a), ridged substrates that produced these boundary layer features received more than 10-fold greater settlement of two species of Caribbean broadcast-spawning coral larvae than compared to flat substrates. The underlying settlement mechanisms were elucidated by computational fluid dynamics coupled with an agent-based model of swimming larvae. These simulations demonstrated that the unique BBLs generated over ridged surfaces direct larvae toward substrates and increase the opportunity for settlement by allowing larvae to be active for longer periods of time than compared to over substrates without topographical features at this length scale. Our findings suggest that while ecologically relevant flows generally make settlement more difficult, the presence of millimeter-scale topography can enable settlement in an otherwise unfavorable hydrodynamic environment.

## Results

**Settlement of coral larvae onto engineered substrates in oscillatory flow.** Larvae of the corals *Diploria labyrinthiformis* (Grooved Brain Coral) and *Colpophyllia natans* (Boulder Brain Coral) were subject to static conditions and controlled oscillatory flow over engineered settlement substrates with either flat surfaces or millimeter-scale ridges (Fig. 1b and 1c). Both flow and substrate topography had significant effects on larval settlement (Fig. 1d). Focusing first on *D. labyrinthiformis* larvae, settlement was 4.3 times lower in oscillatory flow compared to static conditions ($p = 0.0002$, F = 34.5). However, in both flow and static conditions, 17 and 3.8 times more larvae settled onto the ridged substrates than onto the flat substrates, respectively (Fig. 1d; $p = 0.0001$, F = 42.2). Further, on flat substrates in oscillatory flow, settlement was nearly zero (0.33 settlers per run, $n = 3$; Fig. 1d).

The importance of substrate topography in enabling larval settlement was also illustrated by the settlement locations of individual *D. labyrithiformis* larvae (Fig. 1e). Settlement locations in static conditions ($n = 40$ settlers) and oscillatory flow ($n = 11$ settlers) were mapped onto a single section of a ridged substrate for comparison. In both cases, >87% of larvae settled between the ridges rather than on the tops of the ridges. However, settlement patterns in the spaces between the ridges were markedly different depending on flow. In static conditions, larvae settled uniformly across these areas (Fig. 1e, *upper panel*) while in oscillatory flow, 90% of larvae settled within 1 mm of the base of the ridges (Fig. 1e, *lower panel*). In particular, settlement was concentrated near the ridges that were perpendicular to the oscillatory flow.

An unexpected result further highlighted the importance of substrate topography for larval settlement: 67% of *D. labyrinthiformis* larvae and 74% of *C. natans* larvae settled not onto the tops of substrates, but within the cryptic spaces that were formed by gaps between the 3D-printed holding rack and the substrates (due to slight size differences created during fabrication; Fig. 1f, Supplementary Fig. S1). In particular, *C. natans* larvae settled into these cryptic spaces in such large numbers that no significant settlement differences were observed on the tops of the substrates. Overall, we found substantial evidence that substrate topography drives coral larval settlement, especially in wave-dominated, oscillatory flow, regardless of whether the topography was the ridges on the substrates or the gaps between the substrates and their holding rack. Based on these striking results, we conducted a detailed investigation of the boundary layer flow in the flume to visualize and quantify the hydrodynamic mechanisms underlying the preferential settlement of larvae onto ridged substrates and into cryptic habitats.

**Ridged substrates create recirculating boundary layer flow.** Qualitative differences in the flow fields over both substrate types were visualized by tracking fluorescent tracer particles over a full oscillatory period and plotting the resulting data for two characteristic phases of the flow period: peak rightward flow ("peak flow", Fig. 1b, *straight line region*) and the turning point from rightward to leftward flow ("turning point flow", Fig. 1b, *curved line region*). Due to the symmetry of the sinusoidal flow and the substrate rack, these phases are symmetric with the peak leftward phase and the opposing turning point, respectively.

Boundary layer flows differed according to substrate topography. During the peak flow phase, particle pathlines revealed that the bulk flow was laminar far above both substrate types (>>1 mm; Fig. 1f) and, for flat substrates, the boundary layer flow was also laminar near the substrates (<1 mm). However, for ridged substrates, particle pathlines revealed regions of fluid recirculation between the ridges (Fig. 1f, *yellow arrow*). During the turning point phase, regions of recirculation also formed above the gaps between the 3D-printed rack and the sides of the substrates (Supplementary Videos S1 and S2), but no recirculation was observed directly above the flat substrates.

**Topography-induced flow recirculation corresponds to coral settlement location.** Quantitative measurements of boundary layer flows (rotation, strain, and velocity) were obtained using particle tracking velocimetry (PTV) of the fluorescent tracer particles and these data were examined in relation to observed patterns of larval settlement. First, regions of recirculation near substrate surfaces were identified by locating vortex cores in instantaneous velocity fields using the $Q$-criterion metric [38], [39]. The $Q$-criterion is defined as $Q = 1/2(|\Omega|^2 - |S|^2)$, where $\Omega$ is the vorticity tensor and $S$ is the rate-of-strain tensor. A positive value of $Q$ corresponds to a region where the local rotation exceeds the local strain, a characteristic of flow recirculation. A negative value of $Q$ corresponds to a region in the flow field where the local strain exceeds the local rotation, such as in laminar shear flow over a flat surface. However, in practice, a non-zero threshold, $Q_{thresh}$, must be used to accurately identify true recirculation. Here, $Q_{thresh}$ is defined as the standard deviation (SD) in $Q$ to account for spurious experimental results, often encountered when tracking particles near surfaces.

For flat substrates, no recirculation was observed during the peak flow phase (Fig. 2a, *top*). In contrast, for ridged substrates, regions of flow recirculation with $Q$-criteria greater than $Q_{thresh}$ were observed between the ridges. Vortex structures were clearly seen forming and detaching from the trailing edges of each ridge (Supplementary Video S3). At the turning point of the oscillatory period, the reversal of fluid motion produced small, short-lived regions of rotation over flat substrates. However, much greater recirculation was observed during the same phase over the ridged substrates (Fig. 2a, *bottom*).

Local substrate topography affected the maximum $Q$-criterion value that occurred during the flow oscillation period (Fig. 2b). Above flat substrates, the $Q$-criterion always remained lower than the threshold value, indicating that these regions were dominated by fluid strain. However, above ridged substrates, $Q$-criterion values were always much larger than the threshold value, indicating that these regions were dominated by flow recirculation. Here, $Q$-criterion values ranged from 65

s$^{-2}$ at the turning point up to 130 s$^{-2}$ during peak flow (Fig. 2b). *Q*-criterion values greater than $Q_{thresh}$ were also observed above the gaps between the substrates and the 3D-printed rack (Supplementary Fig. S2); here, the values ranged from 40 s$^{-2}$ up to 70 s$^{-2}$ for flat and ridged substrates, respectively (Fig. 2b).

Remarkably, when comparing all settlement locations of both *D. labyrinthiformis* and *C. natans* larvae in the context of the *Q*-criterion data, larval settlement was overwhelmingly concentrated in locations where the *Q*-criterion was greater than $Q_{thresh}$ at some interval during the flow period (Fig. 2c). Indeed, when all larval settlement locations from both species were analyzed according to their corresponding *Q*-criterion, >98% of *D. labyrinthiformis* larvae (*n* = 64) and >91% of *C. natans* larvae (*n* = 33) settled in a region with a *Q*-criterion > $Q_{thresh}$ ($p$ = 0.0013, F = 18.5; Fig. 2c).

**Substrate topography extends the settling window near substrates.** In addition to modifying recirculation in the boundary layer, substrate topography also directly influenced fluid speed near the substrate surfaces (Fig. 3a). Fluid speeds were extracted from PTV data and compared to the average swimming speed of coral larvae to reveal regions in the flow where larvae would either be expected to undergo passive transport or would be capable of swimming against the flow. We used a species-averaged larval swimming speed ($|u_\ell|$) of 3 ± 1 mm s$^{-1}$, which was computed using PTV data obtained from swimming *D. labyrinthiformis* larvae and previously-reported data on larval swimming speeds (Supplementary Table S1) [33].

During the peak flow phase, in regions ≥3 mm above both substrate types, the local flow speed ($|U|$) reached up to 60 mm s$^{-1}$, i.e., up to 20 times faster than the average larval swimming speed ($|u_\ell|$, Fig. 3a, *top*). The flow speed generally decreased closer to the substrates, but the thickness of these lower velocity regions varied between substrate types. For ridged substrates, the average local flow speeds in regions up to 3 mm above the substrate were much lower than in the free stream above, ranging from less than 1 to 10 times the larval swimming speed. In contrast, this low velocity region was much thinner over flat substrates, extending only up to 0.75 mm above the substrate.

During the turning point flow phase, regions far above (≥5 mm) the surfaces of both substrate types also had fluid speeds equal to or less than the average larval swimming speed (i.e, $|U| \leq |u_\ell|$). Closer to the substrate surface (≤1.5 mm), the turning point flow speed was slightly increased over ridged substrates and remained low over flat substrates. However, due to the rapid fluid acceleration before and after the turning point, these regions of low flow speed were transient and short-lived.

Not only was fluid speed near the substrates highly variable during an oscillation period, this variation was strongly influenced by substrate topography (Fig. 3b). The fluid speed in the region

≤1.5 mm above the substrate, which for coral larvae is approximately 3 to 5 body lengths, is considered to be crucial for larval settlement [40]. For flat substrates, the flow speed in this region varied by an order of magnitude, ranging from 1.5 to 15 mm s$^{-1}$ during an oscillation period, or 0.5 to 5 times the average larval swimming speed, with the lowest flow speed occurring during the turning point phase. For ridged substrates, the flow speed in this region ranged from 1.2 to 6 mm s$^{-1}$, or 0.4 to 2 times the average larval swimming speed, with the lowest flow speed occurring during the peak flow phase.

To further illustrate the effects of substrate topography on the dynamics of local flow speed, and consequent opportunities for larval settlement, we quantified the lengths of the larval settling windows that occurred in the regions ≤1.5 mm above each substrate type (Fig. 3b, highlighted regions). Settling windows were defined as time intervals during which $|U|$ dropped below $|u_\ell|$ so that larvae could actively swim toward and attach to a substrate [41]. The flat substrates produced only two brief settling windows, each lasting 0.6 s or ~11% of an oscillatory period, at the turning points of the flow (Fig. 3b). Ridged substrates, however, produced settling windows that were ~3 times longer (1.86 s each). In fact, in the regions between the substrate ridges, the two settling windows comprised the majority of the flow period (~69% or 3.7 s). In addition to producing larger regions of low flow speed and longer settling windows, the ridged substrates also created small, permanent regions of low velocity at the base of each ridge (Fig. 3a, *black arrows*). Even during intervals of high relative flow speed, which did not satisfy the condition of a settling window, the local fluid velocity near the base of the ridges remained equal to or less than the average larval swimming speed. Notably, these regions corresponded to the locations where larval settlement was the highest in the oscillatory flow experiments (Fig. 1e).

**Agent-based simulations of larval transport and settlement further demonstrate the effects of substrate topography.** In agreement with flume settlement experiments, agent-based larval simulations showed much higher rates of settlement onto ridged substrates than flat substrates (Fig. 4a). 79% of larval settlement occurred on ridged substrates compared to only 11% on flat substrates. To probe the sensitivity of this phenomenon, an additional simulation was run using substrates with ridges that were 0.25 mm tall and spaced 0.5 mm apart, length scales similar to the larval length scale and would be expected to increase settlement according to Attachment Point Theory. Surprisingly, larval settlement in these simulations was nearly identical to in simulations of flat substrates, with the 0.25 mm ridges receiving only ~10% of the larvae that settled (Fig. 4a). These striking results prompted us to further investigate the specific trajectories of simulated larvae and the local hydrodynamic environments that produced much greater rates of settlement onto the 2.5 mm ridged substrates in comparison to the substrates that received far fewer settlers.

Larval transport and settlement over 2.5 mm ridged substrates is illustrated by the representative trajectory of a single simulated larva shown in Figure 4b. Initially, this simulated larva was seeded in the flow 1 cm above the substrate (Fig. 4b(i)). The larva was then transported by the oscillatory flow, traveling large distances horizontally while slowly approaching the substrate surface (Fig. 4b(ii - vi)). As the larva neared the substrate, it was directed toward the surface over a ridge (Fig. 4b(viii)). Finally, the larva contacted the substrate surface with a net speed that was lower than the larval swimming speed plus one standard deviation, and the larvae was therefore considered to be settled (Fig. 4b(ix)). Upon closer inspection of the trajectory before settlement, the larva traversed through a region of high rotation in which the $Q$-criterion was above the threshold value, $Q_{thresh}$ (Fig. 4b, zoomed pane). Immediately following, the larva was transported from a high-velocity flow region of $U > 5u_\ell$ to a low-velocity flow region of $U < 5u_\ell$ between the ridges.

The hydrodynamic mechanisms that allowed larval settlement were further elucidated by computing the instantaneous $Q$-criterion and relative swimming speed experienced by the simulate larva along its trajectory. The instantaneous $Q$-criterion fluctuated rapidly and depended on the location of the larva in the flow with respect to the substrate surface (Fig. 4c, *top*). In the bulk flow region, velocity gradients were minimal and therefore the larva experienced $Q$-criterion values close to zero. Near the surface or a ridge, the larva experienced large negative (strain) or positive (rotation) $Q$-criterion values, respectively. Finally, very close to the substrate surface, the larva experienced a low $Q$-criterion value, which enabled settlement.

The instantaneous relative flow speed experienced by the larva fluctuated slowly at first due to the length of the oscillatory flow period, and more rapidly as the larva approached the surface (Fig. 4c, *bottom*). Tracking of the flow speed enabled the identification of "larval active windows," defined as intervals during which the larval swimming speed was greater than the local instantaneous fluid speed, i.e., when $U/u_\ell < 1$. In the bulk flow region, the relative flow speed was high ($U/u_\ell \gg 1$) except for short intervals during the flow turning points. Closer to the substrate, the relative flow speed fluctuated rapidly between high and low values, and the larva experienced multiple consecutive active windows upon arriving between the ridges (Fig. 4c, *bottom*, Supplementary Video S4). After experiencing a final active window close to the substrate surface, the larva was able to settle; this occurred when a larval active window coincided with a settling window ≤1.5 mm from the substrate. Conversely, for simulated trajectories over flat substrates and substrates with 0.25 mm ridges, larvae experienced high shear and strain (reorientation) and fewer larval active windows, therefore preventing high levels of settlement (Supplementary Fig. S3, S4, and S5).

## **Discussion**

The importance of hydrodynamics in larval navigation and settlement has been recognized for some time. Studies have shown that changes in wave intensity can induce larvae to swim upward or downward [42] and high shear forces can trigger settlement competency [43], [44]. As well, specific flow regimes can inhibit recruitment because most larvae have a limited ability to swim against high velocity fields [33], [45] or to settle in the presence of strong oscillatory flows [32], [46], which are often found in the benthic boundary layer (BBL) over reef canopies [35], [47]. Despite the known effects of substrate topography on BBL flow, studies of larval settlement related to topography are still often conducted in static conditions. Thus, the direct link between substrate surface topography, BBL flow, and coral larval settlement remains elusive, yet it is paramount to improve the effectiveness of reef restoration and recovery efforts.

We addressed this knowledge gap by investigating coral larval settlement on engineered restoration substrates under controlled oscillatory flow. Experiments with *D. labyrinthiformis* and *C. natans* larvae revealed two ways in which BBL flows over millimeter-scale ridges enhanced settlement compared to flat surfaces. First, the interaction of the flow and the substrate ridges produced recirculation zones that directed larvae from regions of high velocity – in which they are effectively passive particles – to regions of low velocity where they could actively move in relation to the surrounding fluid. Here, the utilization of particle tracking methods and the *Q*-criterion metric of vortex identification allowed us to locate and quantify transient regions of recirculation that formed over larval settlement locations (Fig. 2). Second, millimeter-scale substrate ridges extended the settling windows in the near-surface flows by keeping velocities equal to or less than the average coral larval swimming speed. The duration of these settlement windows for ridged substrates was extended by at least 300% compared to flat substrates (Fig. 3). These two experimental observations were reinforced by an agent-based simulation, which demonstrated that the combination of fluid recirculation and extended settlement windows over ridged substrates created more opportunities for coral larvae to actively navigate toward and settle onto substrate surfaces (Fig. 4).

Our combined experimental and modeling results illustrate the importance of often-overlooked, millimeter-length substrate features in marine larval settlement. Most reef-scale bathymetric mapping does not quantify topography below a resolution of ~10 cm [48]–[50], while most topography-based settlement studies have focused on the effect of micro-scale features of <1 mm [51], [52]. According to Attachment Point Theory, larvae generally settle within topographical features close to their size (~0.1 – 1 mm) [24]. However, because most controlled larval settlement studies have been performed in static or low-flow conditions [53], they do not fully mimic the complex physical environment in which coral larvae must settle in nature. In our simulations, larval settlement onto surfaces with 0.25 mm tall ridges was not statistically different than settlement onto flat surfaces, and settlement to both of these topography types was low. We did not detect sustained periods comprising any of the key hydrodynamic characteristics associated with increased settlement (e.g., vorticity, rotation) over either of these substrate types in our flume experiments. Therefore, while sub-millimeter-scale topography helps promote larval settlement in low-flow and static conditions, it is unlikely to do so in most wave-dominated, natural reef environments.

To date, coral reef restoration has not been heavily focused on promoting natural coral settlement – particularly in strongly wave-dominated conditions – but this may be possible in the future, especially by building upon and leveraging the increasing diversity of established methods for fostering larval settlement. Traditionally, most laboratory-reared coral juveniles were pre-settled onto substrates under low-flow or static conditions. More recently, restoration approaches have expanded to include direct larval seeding [54] and *in situ* settlement pools [9], which both achieve larval settlement under natural flow regimes that have been dampened somewhat by the seeding/settlement structures themselves. These approaches have revealed essential details about the materials, substrate communities, and benthic communities that allow settlement and survival. Now, insights and innovations gained from these approaches can be combined with hydrodynamic modeling to engineer high-performance substrates designed to attract and entrain larvae, foster settlement, shelter early post-settlement juveniles, promote calcification, and support the dominance of corals relative to other benthic competitors, even in wave-dominated, nearshore environments. Some progress has already been made in this area, e.g., the installation of topographically complex concretes to promote recruitment to seawalls and breakwaters [55]. Nevertheless, facilitating robust coral settlement from natural larval pools remains a growth opportunity for materials engineering and fluid physics.

Here, we focused on engineered substrates and their potential applications in coral reef restoration, but our results also have relevance to the failure of natural coral recruitment that has been observed worldwide [1]–[3]. It is widely accepted that coral reefs have undergone a topographic flattening at the centimeter to meter scale [56]–[58]. Similar, but less appreciated, is the flattening of coral reefs that has occurred at the millimeter to centimeter scale as a consequence of coral loss and algal overgrowth. The majority of now-dominant benthic groups (e.g., turf algae, macroalgae, sponges, and cyanobacteria) form little to no rigid structure at these scales. Yet, these are precisely the type of topographic structures that facilitate coral settlement, millimeter-scale structures that were once formed by bare coral skeletons, CCAs, bivalves, and intense parrotfish and urchin grazing on hard substrates. Here, we have shown that the loss of such topography has a profound effect on the hydrodynamics of larvae settlement, and hence, the restoration of such topography may help to reverse declining recruitment rates and restore them to historical levels.

**Materials and Methods**

**Gamete Collection and Larval Rearing.** Gametes were collected from the hermaphroditic Caribbean broadcast-spawning corals *Diploria labyrinthiformis* (Grooved Brain Coral) and *Colpophyllia natans* (Boulder Brain Coral) at Playa Zakitó, Curaçao (also known as Water Factory; 12°6'34" N, 68°57'18" W). Egg-sperm bundles were collected from 7 *D. labyrinthiformis* colonies and 4 *C. natans* colonies at a depth of 5 – 10 m. Larvae were reared following previously-published methods [59]–[62] which are also summarized in the Supplementary Information. All larval rearing steps and experiments were performed with 0.5 µm filtered seawater (FSW; spun polypropylene sediment filters, sequential pore sizes of 50 µm, 20 µm, 5 µm, and 0.5 µm, $H_2O$ Distributors, Marietta, GA).

**Flume Tank Design and Operation.** Larval settlement experiments and flow visualization measurements were conducted in a custom-built, U-shaped flume tank (Fig. 1a) with the following components: (1) laser sheet, (2) acrylic viewing section, (3) PVC T-socket, (4) custom 3D-printed PVC-to-acrylic connectors, (5) PVC elbow, (6) motor and piston assembly, and (7) high-speed camera. All experiments were run using a sinusoidal oscillatory flow with a period between 5 – 6 s and a peak mean velocity between 3 – 5 cm $s^{-1}$ (Fig. 1b); these parameters were selected to mimic the wave-dominated conditions often encountered in shallow coral reef habitats [33], [46]. Additional information on flume construction and operation is provided in the Supplementary Information.

**Settlement Substrate Fabrication.** Surface topography was controlled by manufacturing calcium carbonate ($CaCO_3$) settlement substrates from an un-aged lime mortar (see Supplementary Information) [63]. Substrates were prepared with either a flat top surface or with ridges that were 2.5 mm tall and spaced 5 mm apart edge-to-edge (Fig. 1c). The final surface topography of both substrate types was characterized with a 3D laser scanning confocal profilometer (Keyence VK-X1000; Fig. 1c). In addition to the presence or absence of millimeter-scale ridged features, both substrate types had inherent microscale topography with a mean roughness of $1.88 \pm 0.23$ µm and a maximum peak height of $13.20 \pm 1.81$ µm (mean ± standard deviation; $n = 3$). For larval settlement experiments, a 3D-printed rack was designed to hold eight individual settlement substrates in the transparent acrylic section of the flume. Substrates were held in a $2 \times 4$ array, with a row of four flat substrates and four ridged substrates each oriented parallel to the oscillating flow (Supplementary Fig. S6).

**Flow Visualization.** The flow fields generated within the flume and over the substrates were characterized by particle tracking velocimetry (PTV). A laser pointer (<5 mW; 405 nm wavelength) was positioned above the transparent acrylic section of the flume and directed through a plano-concave cylindrical lens to create a vertical laser sheet along its length (Fig. 1a). Fluorescent green polyethylene microspheres (Cospheric, 250-300 µm diameter, 0.99-1.01 g cm⁻ [3]) were used as tracer particles. A scientific CMOS camera (Ximea) with a Nikkor macro lens (Nikon) was positioned perpendicular to the laser sheet to capture the movement of the particles over time. Videos were recorded at 90 frames s$^{-1}$ with a spatial resolution of 31 µm per pixel. For each flow phase (i.e., *peak* flow and *turning point* flow), PTV data were averaged over 40 frames collected across the respective phase ($\Delta t = 0.44$ s). Representative videos of the flow over flat and ridged substrates in the 3D-printed rack in the flume can be found in the Supplementary Information (Video S1 and S2). A separate, 3D-printed substrate with features matching the 2D profile of the ridged $CaCO_3$ substrates was used to visualize the flow field between each millimeter-scale ridge (Supplementary Fig. S7 and Video S5).

**Larval Settlement Experiments.** For all settlement experiments, 750 – 1000 larvae were added into the flume tank at a density of 50 – 70 larvae L$^{-1}$ FSW. The motor drive and piston were set to produce a flow speed (3 – 5 cm s$^{-1}$) and period (5 – 6 s) that mimicked observed reef conditions. Control replicates without an applied flow (i.e., static conditions) were also run. Experiments were scored after 1 day for *D. labyrinthiformis* and after 2 days for *C. natans* (because the latter were slower to settle). During scoring, the location (substrate top, side, bottom, or rack) and number of settlers was recorded for each substrate type (Supplementary Fig. 1). *D. labyrinthiformis* experiments were repeated three times ($n = 3$), and *C. natans* experiments were repeated four times ($n = 4$) for both flow and static conditions. Additionally, the exact locations of settlers were mapped for two *D. labyrinthiformis* runs. To account for random effects related to substrate position, the flume was kept in a temperature-controlled room and the location of the flume within the room and the placement of the flat and ridged substrates were alternated for each replicate run.

**Agent-Based Simulation of Larval Settlement.** A finite-element model of the flume was developed in COMSOL Multiphysics software to simulate the boundary layer velocity fields that formed over different substrate topographies under an applied oscillatory period of 5.5 s and a mean free stream velocity of 4.5 cm s$^{-1}$. The flow simulation was validated for mesh convergence and periodicity (Fig. S8 and S9) and the resultant velocity fields were imported into MATLAB to compute larval trajectories in the boundary layer flows that developed. Larvae were modeled as neutrally-buoyant ellipses with a constant swimming speed along the direction of their major axis. Their translational motion was calculated using the equation:

$$\dot{r} = U + u_\ell \hat{n} \qquad (1)$$

where $\dot{r}$ is the total larval velocity, $U$ is the local flow velocity, $u_\ell$ is the larval swimming speed, and $\hat{n}$ is the orientation of the larva's major axis. Their rotational motion was calculated using the equation:

$$\dot{\theta} = \frac{\omega_z}{2} + \alpha \hat{g} \cdot E \hat{n} \qquad (2)$$

where $\dot{\theta}$ is the total larval angular velocity, $\omega_z$ is the vorticity, $\alpha$ is the shape parameter, $E$ is the symmetric rate of strain tensor, and $\hat{g}$ is the unit vector along the larva's minor axis. The shape parameter is defined as: $\alpha = (1 - (b/a)^2)/(1 + (b/a)^2)$, where $a$ is the semi-major axis and $b$ is the semi-minor axis. We simulated larvae with $a = 0.25$ mm, $b = 0.15$ mm, and a swimming speed of 3 mm s$^{-1}$. Trajectories were computed until either (1) larvae encountered a surface while their total speed ($\dot{r}$) was lower than the larval swimming speed plus one standard deviation, at which time they were considered to be settled on the substrate, or (2) until larvae exited the top of the simulation domain.

**Statistical Analysis.** Significant differences in settlement between substrate types under different flow conditions were determined with two-way analysis of variance (ANOVA) and post-hoc Tukey's Honestly Significant Difference (HSD) tests using a significance level ($\alpha$) of 0.05. The number of settled larvae in a particular location was converted into a proportion of the total larvae added ($n = 1000$) and normalized using an angular transformation (arcsine square root) prior to significance testing [64]. The flow condition (static *versus* flow) and substrate topography (flat *versus* ridged) were assumed to be fixed variables, and the test had 11 degrees of freedom (df). For both species, data from larval settlement onto the substrates and into the gap regions were also recast according to the calculated *Q*-criterion value above each location [64]. Here, the *Q*-criterion (below or above $Q_{thresh}$) and species (*D. labyrinthiformis* or *C. natans*) were treated as fixed factors for two-way ANOVA (df = 13). Settlement events were assumed to have a negligible effect on the supply of available larvae because less than 10% of all larvae settled in each experimental run. Differences in settlement in the agent-based larval simulations were analyzed with one-way ANOVA and post-hoc Tukey HSD tests assuming substrate topography to be a fixed variable (df = 239).

**Acknowledgements**

We thank the National Science Foundation for funding under the Convergence RAISE program (#IOS-1848671). Field research was conducted with the support of CARMABI Foundation staff and with permits from the Government of Curaçao Ministry of Health, Environment, and Nature (GMN). Characterization of settlement substrates was carried out in part in the Materials Research Laboratory Central Research Facilities, University of Illinois. Additional support for coral spawning fieldwork in 2019 was provided by the Paul G. Allen Family Foundation (to K.L.M.). The authors thank V. Chamberland and K. Latijnhouwers of SECORE International for their logistical support and their assistance with collection and care of *C. natans* larvae. We also thank the many students, interns, and volunteers at CARMABI during the 2019 coral spawning season including M.-J. Bennett, E. Culbertson, T. Doblado Speck, D. Flores, M. Ramirez, Z. Ransom, S. Schönherr, N. Le Trocquer, and the staff of The Diveshop Curaçao.

**Figures**

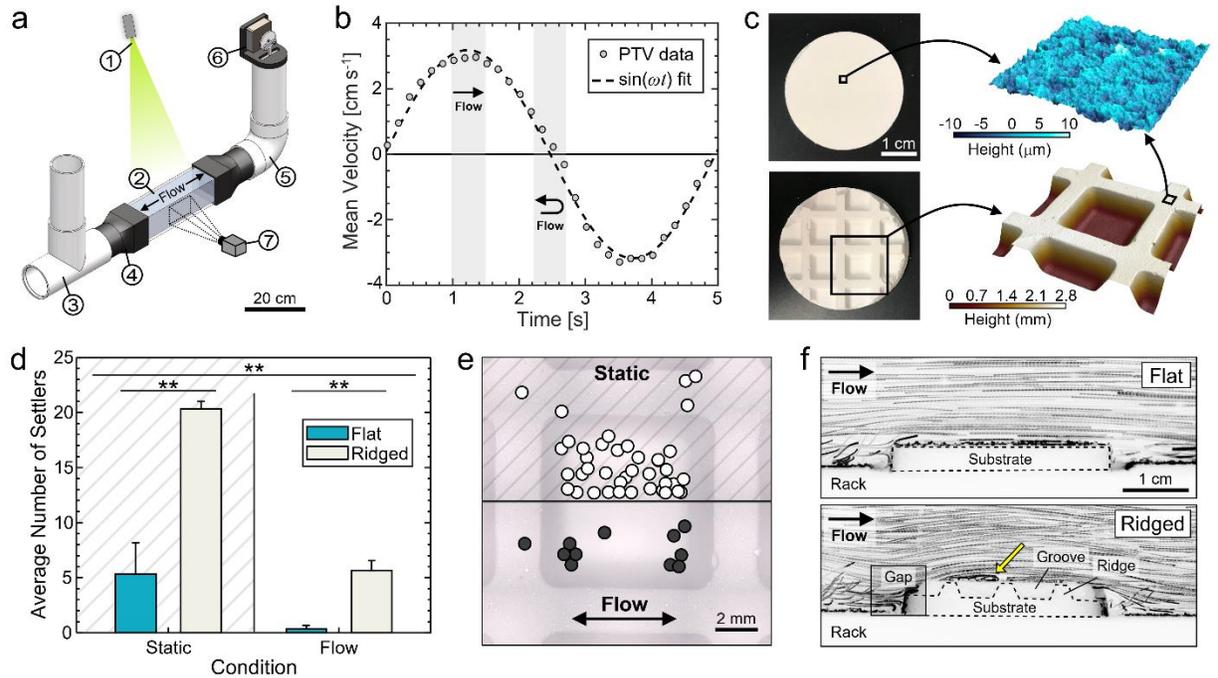

**Figure 1**. Settlement of coral larvae onto flat and ridged substrates in oscillatory flow. (a) Schematic of the oscillatory flume tank and particle tracking system. Major components include (1) laser sheet, (2) acrylic viewing section, (3) PVC T-socket, (4) custom 3D-printed PVC-to-acrylic connectors, (5) PVC elbow, (6) motor and piston assembly, and (7) high-speed camera. (b) Spatially-averaged fluid velocity in the central viewing section of the flume obtained from particle tracking velocimetry measurements (PTV; circles) over a full oscillation period. The grey highlighted sections correspond to the phases of peak flow (*straight arrow*) from left to right and the turning point (*curved arrow*) from rightward to leftward flow. (c) Photographs of the flat and ridged $CaCO_3$-based settlement substrates and 3D laser confocal maps showing the micro-scale topography of both substrate types (*top*) and the millimeter-scale ridges of the ridged substrates (*bottom*). (d) Larval settlement data on flat and ridged substrates in static and oscillatory flow conditions (\*\*: $p < 0.001$; post-hoc Tukey HSD). (e) Top view of a ridged substrate illustrating larval settlement locations in static (*top*, $n = 40$) and flow (*bottom*, $n = 11$) conditions overlaid onto a single ridged section. (e) Tracer particle pathlines during peak flow above flat (*top*) and ridged substrates (*bottom*). A region of flow recirculation above the ridged substrate is identified (*yellow arrow*).

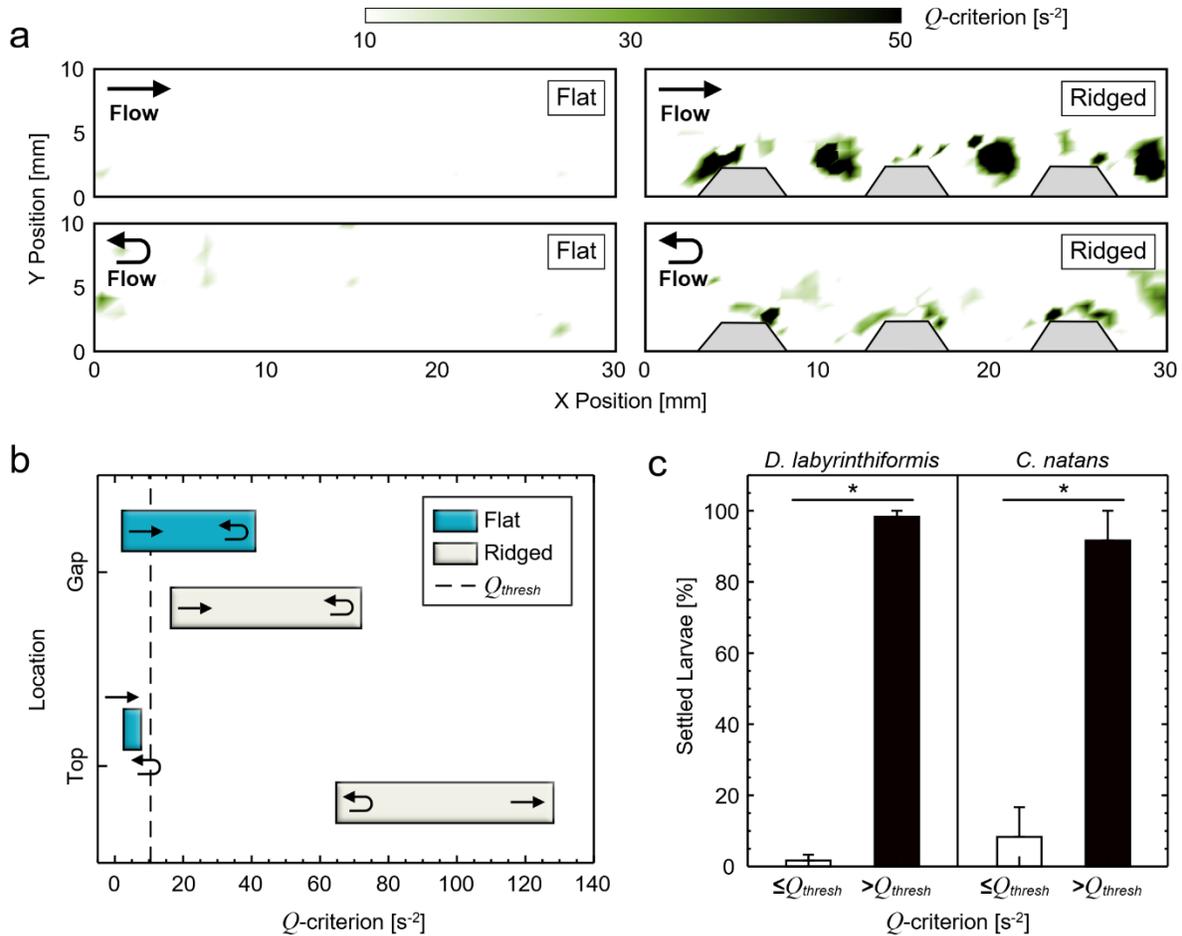

**Figure 2**. Millimeter-scale ridges generate regions of flow recirculation that influence larval settlement. (a) Regions of flow recirculation identified over flat (*left*) and ridged (*right*) substrates during peak (*top*) and turning point (*bottom*) flow using the *Q*-criterion metric. (b) Range plot of the maximum *Q*-criterion values over flat and ridged substrates and gap regions during a full period of oscillation. The dotted line is the average standard deviation in *Q*-criterion across all measurements, which was used as the threshold for the identification of vortex structures ($Q_{thresh}$). (c) Larval settlement data under a flow region with a *Q*-criterion greater than or less than $Q_{thresh}$ for *D. labyrinthiformis* (*left*) and *C. natans* (*right*). Settlement is presented as the mean percent of total settlers in each experimental run and the error bars represent the standard error of the mean. There was a significant difference in the settlement between regions above and below the $Q_{thresh}$ for both coral species (*: $p < 0.01$; post-hoc Tukey HSD).

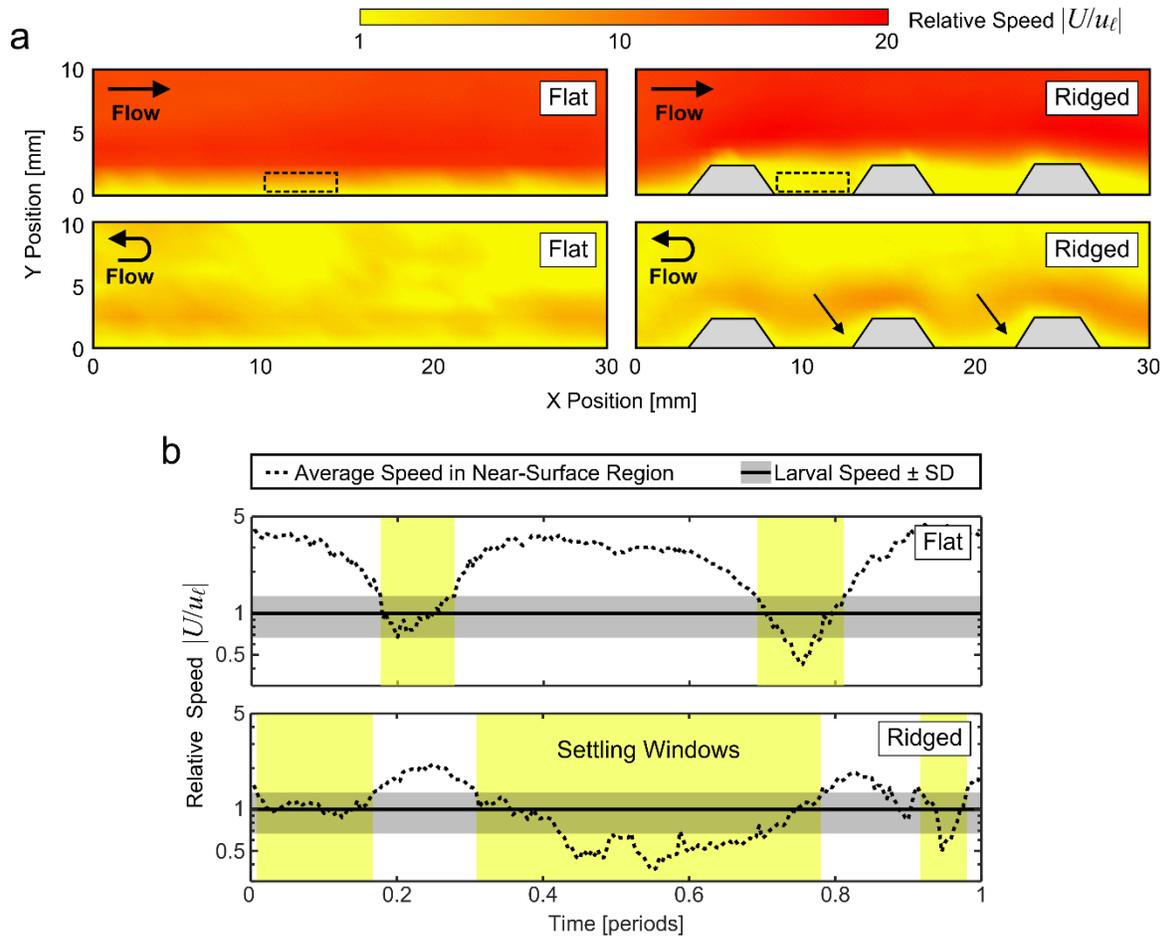

**Figure 3**. Millimeter-scale ridges increase the settling window duration. (a) Visualization of the relative fluid speed over flat (*left*) and ridged (*right*) substrates during peak (*top*) and turning point (*bottom*) flow. The fluid velocity (*U*) is normalized by the average larval swimming velocity ($u_\ell$). Dotted black boxes represent regions ≤1.5 mm above the substrate surface that were used to calculate settlement windows. Black arrows near the bottoms of the ridges indicate regions where the velocity remains low even at the turning points. (b) The average relative flow speed within the dotted black regions plotted over an average period for flat (*top*) and ridged (*bottom*) substrates. The yellow regions highlight the settling windows during which the local flow speed drops below $u_\ell$ (black line with grey band showing its standard deviation).

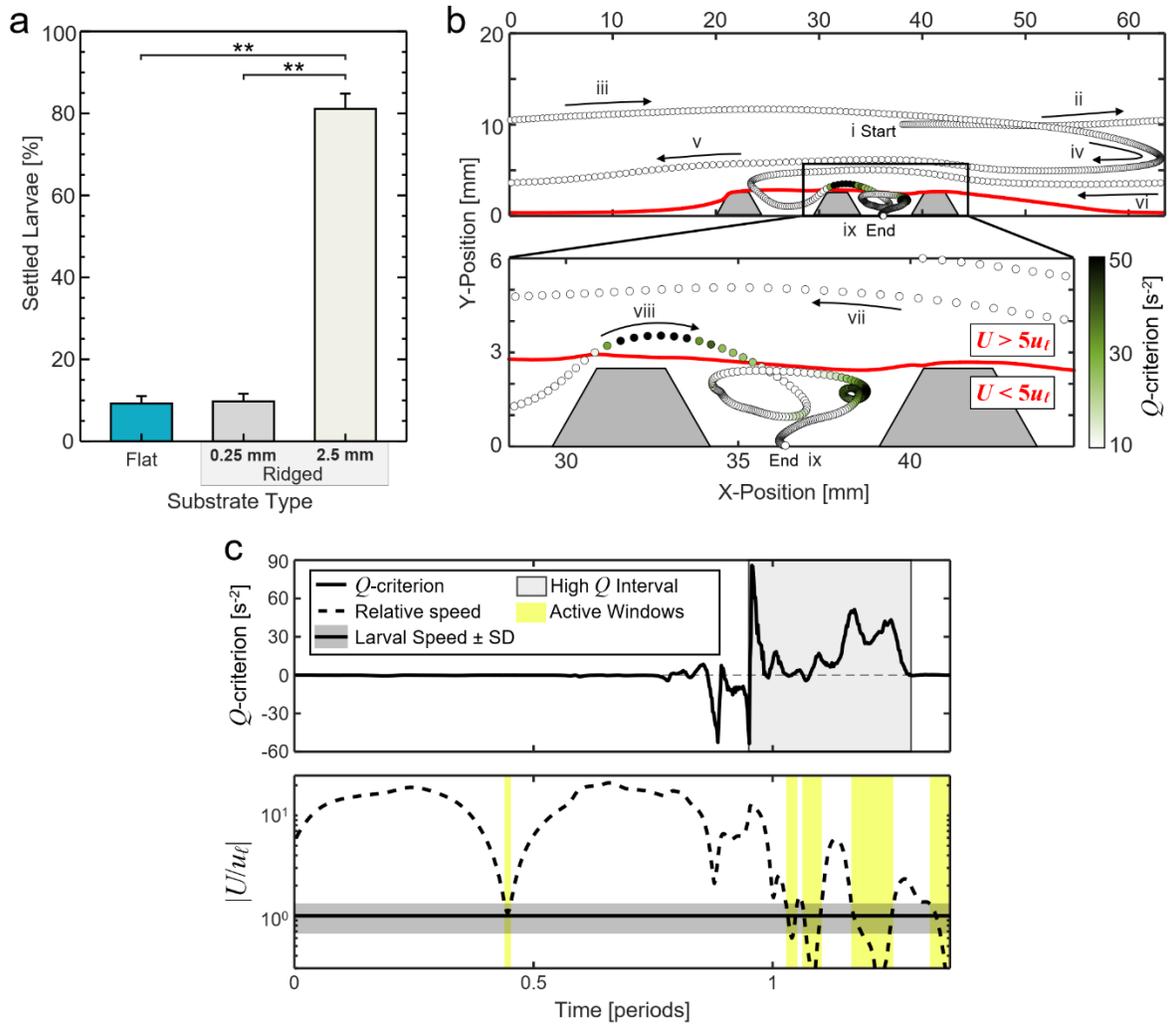

**Figure 4**. Millimeter-scale ridges modify the boundary layer flow to facilitate larval transport and settlement. (a) Results of larval settlement simulations on flat substrates and ridged substrates with millimeter-scale (2.5 mm) and sub-millimeter-scale (0.25 mm) ridges in oscillatory flow. There were significant differences in settlement between substrates with millimeter-scale ridges and flat and sub-millimeter-scale ridged substrates (**: $p < 0.001$; post-hoc Tukey HSD). (b) Simulation of a larva trajectory (colored dots) over a surface with millimeter-scale ridges. The color of the dot indicates the instantaneous $Q$-criterion experienced by the larva at each location. The spacing between dots increases with larval speed. The red line shows the boundary between regions of $U > 5u_\ell$ and $U < 5u_\ell$ during the peak rightward flow phase. (c) The instantaneous $Q$-criterion (*top*) and relative flow speed (*bottom*) experienced by the simulated larva in (b). After encountering multiple regions with high-$Q$ values (grey band), the larva experiences several intervals of low relative fluid velocity, or active windows (yellow bands), before settling.

21